\documentclass[aps,prl,groupedaddress,twocolumn]{revtex4}
\usepackage{amsmath,amsfonts,amssymb,graphics,graphicx,epsfig,color,times,bbm,natbib}

\newcommand{\be}{\begin{equation}}
\newcommand{\ee}{\end{equation}}
\newcommand{\ba}{\begin{eqnarray}}
\newcommand{\ea}{\end{eqnarray}}
\newcommand{\ban}{\begin{eqnarray*}}
\newcommand{\ean}{\end{eqnarray*}}

\newcommand{\sandwich}[3]{\mbox{$ \langle #1 | #2 | #3 \rangle $}}
\newcommand{\ket}[1]{\mbox{$ | #1 \rangle $}}
\newcommand{\bra}[1]{\mbox{$ \langle #1 | $}}

\newcommand{\si}{\sigma}

\newcommand{\one}{\leavevmode\hbox{\small1\normalsize\kern-.33em1}}

\begin{document}

\title{Detection loophole in asymmetric Bell experiments}
\author{Nicolas Brunner }
\email{nicolas.brunner@physics.unige.ch}
\author{Nicolas
Gisin}
\author{Valerio Scarani}
\author{Christoph Simon}
\address{Group of Applied Physics, University of Geneva, CH-1211 Geneva 4, Switzerland}
\date{\today}

\begin{abstract}
The problem of closing the detection loophole with asymmetric
systems, such as entangled atom-photon pairs, is addressed. We show
that, for the Bell inequality $I_{3322}$, a minimal detection
efficiency of $43\%$ can be tolerated for one of the particles, if
the other one is always detected. We also study the influence of
noise and discuss the prospects of experimental implementation.
\end{abstract}
\maketitle


Non-locality is one of the most striking properties of quantum
mechanics. Two distant observers, each holding half of an entangled
quantum state and performing appropriate measurements, share
correlations which are non-local, in the sense that they violate a
Bell inequality \cite{Bell64}. In other words, those correlations
cannot be reproduced by any local hidden variable (lhv) model. All
laboratory experiments to date have confirmed quantum non-locality
\cite{Clauser,Aspect,Weihs,Tittel,Rowe}. There is thus strong
evidence that Nature is non-local. However, considering the
importance of such a statement, it is crucial to perform an
experiment free of any loopholes, which is still missing today.
Another motivation comes from quantum information science, where the
security of some quantum communication protocols is based on the
loophole-free violation of Bell inequalities \cite{NLcrypto}.

Performing a loophole-free Bell test is quite challenging. One first
has to ensure that no signal can be transmitted from one particle to
the other during the measurement process. Thus the measurement
choice on one side and the measurement result on the other side
should be spacelike separated. If this is not the case, one particle
could send some information about the measurement setting it
experiences to the other particle. This is the locality loophole
\cite{Speakable}. Secondly the particles must be detected with a
high enough probability. If the detection efficiency is too low, a
lhv model can reproduce the quantum correlations. In this picture a
hidden variable affects the probability that the particle is
detected depending on the measurement setting chosen by the
observer. This is the detection loophole \cite{Pearle}.

In practice, photon experiments have been able to close the locality
loophole \cite{Aspect,Weihs,Tittel}. However the optical detection
efficiencies are still too low to close the detection loophole. For
the Clauser-Horne-Shimony-Holt (CHSH) \cite{CHSH} inequality, an
efficiency larger than $82.8 \% $ is required to close the detection
loophole with maximally entangled states. Surprisingly, Eberhard
\cite{Eberhard} showed that this threshold efficiency can be lowered
to $66.7 \%$ by using non-maximally entangled states. On the other
hand an experiment carried out on trapped ions \cite{Rowe} closed
the detection loophole, but the ions were only a few micrometers
apart. It would already be a significant step forward to close the
detection loophole for well separated systems. Recently new
proposals for closing both loopholes in a single experiment were
reported \cite{Irvine,Cerf}.

In this paper we focus on asymmetric setups, where the two particles
are detected with different probabilities. This is the case e.g. in
an atom-photon system: the atom is measured with an efficiency close
to one while the probability to detect the photon is smaller.
Intuition suggests that if one party can do very efficient
measurements, then the minimal detection efficiency on the other
side should be considerably lowered compared to the case where both
detectors have the same efficiency. Experimentally this approach
might be quite promising, since recent experiments have demonstrated
atom-photon entanglement \cite{Weinfurter,blinov} and violation of
the CHSH inequality \cite{Monroe}. In the following, after
presenting the general approach to the study of the detection
loophole in asymmetric systems, we focus on the case where one of
the systems is detected with efficiency $\eta_A=1$ and we compute
the threshold efficiency $\eta_B^{th}$ for the detection of the
other system. The best results are obtained for the the
three-setting $I_{3322}$ inequality \cite{dan}. In analogy to
Eberhard's result \cite{Eberhard}, we show that non-maximally
entangled states require a lower efficiency; moreover here, the
threshold goes down to $\sim 43\%$. Then we study two noise models:
background noise and noisy detectors. Finally we discuss the
feasibility of experiments in the light of these results.

{\it General approach.} Let us consider a typical Bell test
scenario. Two distant observers, Alice and Bob, share some quantum
state $\rho_{AB}$. Each of them chooses randomly between a set of
measurements (settings) $\{A_{i}\}_{i=1..N_{A}}$ for Alice,
$\{B_{j}\}_{j=1..N_{B}}$ for Bob. The result of the measurement is
noted $a$, $b$. Here we will focus on dichotomic observables
(corresponding to Von Neumann measurements on qubits) and Alice and
Bob will use the same number of settings, i.e. $a,b \in \{0,1\}$ and
$N_{A}=N_{B} \equiv N$. Repeating the experiment many times, the two
parties can determine the joint probabilities $p(a,b|A_{i},B_{j})$
for any pairs of settings, as well as marginal probabilities
$p(a|A_{i})$ and $p(b|B_{j})$. A Bell inequality is a constraint on
those probabilities, which is satisfied for all lhv models. We say
that a quantum state is non-local if and only if there are
measurement settings such that a Bell inequality is violated.
Mathematically speaking a Bell inequality is a polynomial of joint
and marginal probabilities. In the case $N=2$ the only relevant Bell
inequality is the CHSH inequality, which is defined here using the
Clauser-Horne polynomial \cite{CH} \ba\label{CHSH}\nonumber I_{CHSH}
&=& P(A_{1}B_{1})+P(A_{1}B_{2})+P(A_{2}B_{1})
\\  & & -P(A_{2}B_{2})-P(A_{1})-P(B_{1})   \,\, , \ea
where $P(A_{i}B_{j})$ is a shortcut for $P(00|A_{i}B_{j})$, the
probability that $a=b=0$. The bound for lhv models is $I_{CHSH}\leq
0$, while quantum mechanics can reach up to
$I_{CHSH}=\frac{1}{\sqrt{2}}-\frac{1}{2}$. We also introduce the
Bell polynomial \ba\label{I3322}\nonumber I_{3322} &=&
P(A_{1}B_{1})+P(A_{1}B_{2})+P(A_{1}B_{3}) +P(A_{2}B_{1})
\\\nonumber & & P(A_{2}B_{2})+P(A_{3}B_{1})-P(A_{2}B_{3})
-P(A_{3}B_{2}) \\  & & -2P(A_{1})-P(A_{2})-P(B_{1}) \ea which is the
only relevant Bell inequality for the case $N=3$ \cite{dan}. The
local limit is $I_{3322}\leq 0$ and quantum mechanics violates
$I_{3322}$ up to $\frac{1}{4}$.

As an introductory example, consider the case where Alice and Bob
share maximally entangled states and detect their particles with the
same limited efficiency $\eta$; since they must always produce an
outcome, they agree to output ``0'' in case of no detection. When
both detectors fire, $I_{CHSH}\equiv Q=\frac{1}{\sqrt{2}}-
\frac{1}{2}$. When only Alice's detector fires,
$P(A_{i}B_{j})=P(A_{1})=\frac{1}{2}$ while $P(B_{1})=1$, therefore
$I_{CHSH}\equiv M_A=-\frac{1}{2}$; similarly, when only Bob's
detector fires $I_{CHSH}\equiv M_B=-\frac{1}{2}$. When no detector
fires, the lhv bound is reached, $I_{CHSH}=0$. Consequently, the
whole set of data violates the CHSH inequality if and only if \ba
\eta^2 Q + \eta (1-\eta)(M_A+M_B) \geq 0 \, , \ea yielding the
well-known threshold efficiency $\eta \geq 82.84 \%$.

In general, Alice and Bob test an inequality $I\leq L$ on a state
$\rho_{AB}$ having two different detection efficiencies, $\eta_{A}$
and $\eta_{B}$. In analogy to the previous example, Alice and Bob
must choose the measurement settings $\{A_i,B_j\}$ and the local
strategy for the case of no detection, in order to maximize
\ba\label{etaAetaB}\nonumber I_{\eta_{A},\eta_{B}} &=&  \eta_{A}
\eta_{B} Q + \eta_{A}(1-\eta_B) M_{A} +(1-\eta_A)\eta_{B} M_{B} \\
& & + (1-\eta_A)(1-\eta_B) X \quad , \ea where $Q = Tr({\mathcal{I}}
\rho_{AB})$ is the mean value of the Bell operator ${\mathcal{I}}$
associated to the inequality, $M_{A,B}$ and $X$ are the values of
$I$ obtained from the measurements and the local strategies when one
or both detectors don't fire. We stress that the measurement
settings that maximize $I_{\eta_A,\eta_B}$ are {\it not} those that
maximize $Q$ for the same quantum state, except for the maximally
entangled state. Also, it seems intuitively clear that the optimal
local strategies are such that $X=L$, though we have no general
proof of this.

{\it Case study: $\eta_A=1$.} The general approach above can be
carried out for any specific values of the efficiencies; now we
consider the limit where Alice's detector is perfect, $\eta_{A}=1$.
Moreover, we consider inequalities such that $L=0$. From
(\ref{etaAetaB}) one obtains immediately that the efficiency of
Bob's detector must be above the threshold \ba \eta_{B} &>&
\eta_B^{th}\,=\, \frac{1}{1-Q/M_{A}} \ea in order to close the
detection loophole. For any given state, the measurement settings
and Bob's local strategy in case of no detection must be chosen as
to maximize $|Q/M_{A}|$. Note that if $L=0$, $M_A\leq 0$ since the
event where only Alice's detector clicks is indeed local.

\begin{figure}
\includegraphics[width=\columnwidth]{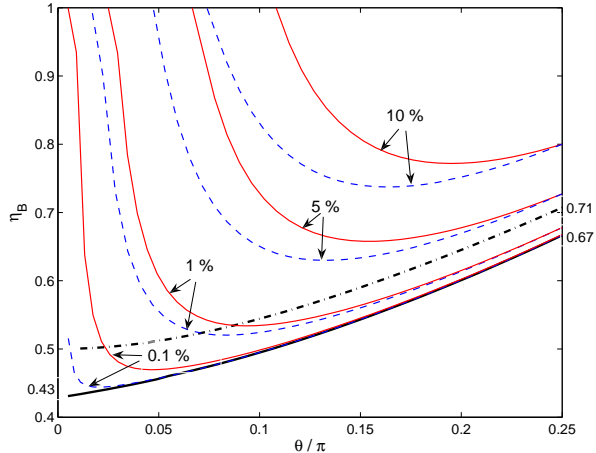} \caption{(Color online)
Numerical optimization of the threshold efficiency $\eta_{B}^{th}$
as a function of $\theta$. Thick lines: pure states result for CHSH
(dashed-dotted line) and $I_{3322}$ (full line). Thin full lines:
$I_{3322}$ for states (\ref{werner}); thin dashed lines: $I_{3322}$
for states (\ref{statedark}); with error value for both.}
\label{noise}
\end{figure}

Consider first pure states. For the {\em maximally entangled state},
one obtains $\eta_B^{th} =\frac{1}{\sqrt{2}}\approx 70.7 \% $ for
the CHSH inequality (the optimal strategy is the same as above) and
$\eta_B^{th} =\frac{2}{3}\approx 66.7 \%$ for the $I_{3322}$
inequality (the settings are those that achieve $Q=\frac{1}{4}$
\cite{dan}, and in the absence of detection Bob outputs ``0''
leading to $M_A=-\frac{1}{2}$). Note that a lhv model is known,
which reproduces the correlations of the maximally entangled state
under the assumption $\eta_A=1$ and $\eta_B=50\%$ \cite{GisinGisin};
it is an interesting open question to close this gap by finding
either a better Bell-type inequality, or a better lhv model. For
{\em pure non-maximally entangled states}
$\ket{\psi_\theta}=\cos\theta \ket{00}+\sin\theta \ket{11}$, we
performed a numerical minimization of $\eta_B^{th}$: we find that
$\eta_B^{th}$ decreases with decreasing $\theta$ both for CHSH and
$I_{3322}$, as shown in Fig.~\ref{noise} (thick lines), in analogy
with Eberhard's result \cite{Eberhard}. The optimal settings can
always be found to lie in the (x,z) plane of the Bloch sphere, i.e.
$ {\mathcal{A}}_{i} =\cos(\alpha_i)\si_z+\sin(\alpha_i)\si_x$ and
${\mathcal{B}}_{j}=\cos(\beta_j)\si_z+\sin(\beta_j)\si_x$. In case
of no detection, we found that it is optimal for Bob to outputs
always ``0''; note that in this case,
$M_A=P(A_1)-1=\frac{1}{2}(\sandwich{\psi_\theta}{{\mathcal{A}}_1
\otimes \one}{\psi_\theta}-1)$ for both inequalities we consider
here, CHSH and $I_{3322}$. In the limit of weakly entangled states,
one finds $\eta_B^{th}\rightarrow 50 \%$ for CHSH and
$\eta_B^{th}\rightarrow \sim 43 \% $ for $I_{3322}$. It is
remarkable that the detection loophole can in principle be closed
with $\eta_B<50 \%$. Though we could not find an analytical
expression for the optimal settings as a function of $\theta$, we
provide a numerical example: for $\theta = \frac{\pi}{100}$,
$I_{3322}$ gives $\eta_B^{th}\simeq 43.3\%$ ($Q\simeq0.0013$ and
$M_A\simeq-0.001$) for the optimal settings $\alpha_0=-0.0012\pi$,
$\alpha_1=0.1331\pi$, $\alpha_2=0.5494 \pi$, $\beta_0=0.0101\pi$,
$\beta_1=-0.0038\pi$ and $\beta_2=-0.0924 \pi$.

\begin{figure}
\includegraphics[width=\columnwidth]{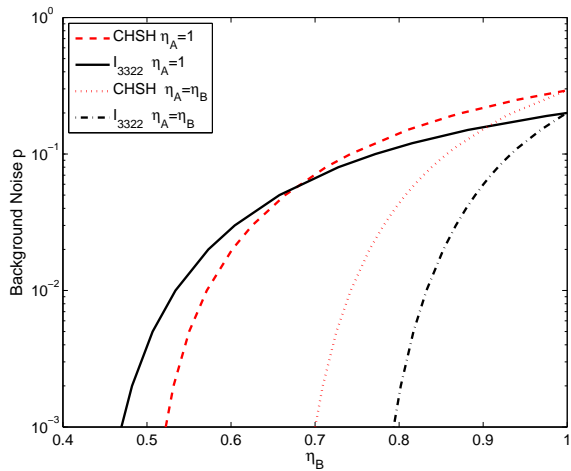} \caption{Minimal detection
efficiency $\eta_{B}$ required for a given noise power. The
curves for the symmetric case, $\eta_{A}=\eta_{B}$, are
also plotted, the curve for CHSH being Eberhard's result.
Though CHSH provides a larger threshold efficiency for any
noise power than $I_{3322}$ in the symmetric case,
$I_{3322}$ can tolerate smaller efficiencies than CHSH when
$p< 6 \%$ and $\eta_{A}=1$. This is probably due to the
fact that inequality $I_{3322}$ is asymmetric, contrary to
CHSH.}\label{noiselog}
\end{figure}

We have seen that $\eta_B^{th}$ decreases with the degree of
entanglement for pure states. However, the violation of the
inequality decreases as well. It is therefore important to study the
effect of noise. We consider two models of noise. The first is
background noise as in Ref.~\cite{Eberhard}: Alice and Bob share a
state of the form \ba\label{werner} \rho_{AB} = (1-p)
\ket{\psi_\theta}\bra{\psi_\theta} + p \frac{\one}{4}\,. \ea For
$\theta = \frac{\pi}{4}$, the state (\ref{werner}) is the Werner
state. The threshold efficiency for $I_{3322}$ as a function of
$\theta$ is shown in Fig.~\ref{noise} (thin full lines). As
expected, when $\theta$ decreases, the threshold efficiency reaches
a minimum: for less entangled states the violation of the inequality
is rapidly overcome by the noise. In Fig.~\ref{noiselog}, one sees
that the $I_{3322}$ inequality can tolerate lower efficiencies than
the CHSH inequality for $p\lesssim 6 \%$.

Another noise model, probably more relevant for experiments,
supposes that Alice's and Bob's detectors have a certain probability
of error, $\varepsilon_d^{A}$ and $\varepsilon_d^{B}$, e.g. due to
dark counts in the case of photon detection. The statistics are then
described by the state \ba\nonumber \rho_{AB} &=&
(1-\varepsilon_d^{A})(1-\varepsilon_d^{B})
\ket{\psi_\theta}\bra{\psi_\theta} \,+ \, \varepsilon_d^{A}
(\frac{\one}{2}\otimes \rho_{B} )
\\ & &+ \varepsilon_d^{B} (\rho_{A} \otimes \frac{\one}{2}  ) \,+\, \varepsilon_d^{A}
\varepsilon_d^{B}\frac{\one}{4}\quad, \label{statedark}\ea where
$\rho_{A}=Tr_{B}\ket{\psi_\theta}\bra{\psi_\theta}$ and
$\rho_{B}=Tr_{A}\ket{\psi_\theta}\bra{\psi_\theta}$ are the reduced
states of Alice and Bob. In the recent atom-photon experiment done
in Munich \cite{Weinfurter}, the atom measurement has $\eta_A\approx
1$ and $\varepsilon_d^{A}\approx 5 \%$, whereas the photon
measurement is much less efficient but also less noisy. In the light
of this, we focus for definiteness on the case $\eta_{A}=1$ and
$\varepsilon_d^{B}=0$. Again, the computed threshold efficiency as a
function of $\theta$ is shown in Fig.~\ref{noise} (thin dashed
lines). The behaviour is qualitatively the same as for the
background noise, but the threshold efficiencies are lower. We have
also found that $I_{3322}$ can tolerate higher error rates than CHSH
as soon as $\eta_{B}<75 \%$. Note that for both noise models, the
optimal settings can be found to lie in the (x,z) plane of the Bloch
sphere and that the optimal strategy for Bob in case of no detection
is to output always ``0''.

{\it Experimental feasibility.} Atom-photon entanglement
has been demonstrated both with Cd ions in an asymmetric
quadrupole trap \cite{Monroe,blinov} and with Rb atoms in
an optical dipole trap \cite{Weinfurter}. Non-maximally
entangled atom-photon states were already created in
Ref.~\cite{blinov}. The overall photon detection efficiency
is very low in these experiments, mostly due to inefficient
photon collection. The collection efficiency could be
brought to the required level by placing the atom inside a
high-finesse cavity. For example, Ref. \cite{mundt}
demonstrated coupling of a trapped ion to a high-finesse
cavity and achieved $\beta=0.51$, where $\beta$ is the
fraction of spontaneously emitted photons that are emitted
into the cavity mode. The experimental conditions in Ref.
\cite{birnbaum} correspond to $\beta$ very close to 1. In
real experiments there are other sources of loss, such as
propagation losses and detector inefficiency. However,
detection efficiencies of order 90 \% have already been
achieved \cite{detectors}, and propagation losses can be
kept small for moderate distances, cf. below. Overall, the
perspective for closing the detection loophole for two
well-separated systems seems excellent using atom-photon
implementations.

Performing a loophole-free Bell experiment requires
enforcing locality of the measurements \cite{Weihs,Tittel}
in addition to closing the detection loophole. The
measurement of the atomic state, which is typically based
on detecting fluorescence from a cycling transition, is
relatively slow. As a consequence, enforcing locality in an
experiment with atom-photon pairs requires a large
separation between the two detection stations for the atom
and the photon. For example, Ref. \cite{Irvine} estimated
that for trapped Ca ions the atomic measurement could be
performed in 30 $\mu$s, assuming that 2\% of the photons
from the cycling transition are collected, leading to a
required separation of order 5 km for an asymmetric
configuration \cite{distance}. For distances of this order
propagation losses for the photon become significant. For
example, 5 km of telecom fiber have a transmission of order
80 \% for the optimal wavelength range around 1.5 $\mu$m,
but only of order 30 \% for wavelengths around 850 nm
\cite{gisincrypto}. Provided that one can achieve fast
atomic measurements, a photon wavelength that minimizes
propagation losses, and highly efficient photon detection,
a loophole-free Bell experiment might be possible with
asymmetric atom-photon systems.

{\it Conclusions and Outlook.} We discussed the detection
loophole in asymmetric Bell tests. In particular we showed
that, for the inequality $I_{3322}$, a minimal detection
efficiency of $\eta_B = 43 \%$ can be tolerated (for
$\eta_A=1$), considering non-maximally entangled states.
For maximally entangled states, the threshold efficiency is
$\eta_B= 66.7\%$. For these states the lhv model of Ref.
\cite{GisinGisin}, based on the detection loophole,
provides a lower bound for the threshold efficiency
$\eta_B>50\%$. It is an interesting question wether this
bound can be reached by considering other Bell
inequalities. We have found no improvement using the
following inequalities: $I_{4422}$ and $I_{3422}^{1,2,3}$
from Ref. \cite{dan}, $A_5$ from Ref. \cite{Avis}, and
$AS_{1,2}$ from Ref. \cite{AS}. From an experimental point
of view, we have argued that atom-photon entanglement seems
promising for closing the detection loophole for well
separated systems. We also briefly discussed the
experimental requirements for realizing a loophole-free
Bell experiment using an asymmetric approach.

 {\it Note added in proof.} While finishing the writing of
this manuscript, we became aware that the results presented here
about the CHSH inequality were independently derived by Cabello and
Larsson \cite{Cabello}.

The authors thank A.A. M\'ethot, J. Volz, H. de Riedmatten and M.
Legr\'e for discussions, and S. Pironio for pointing out a mistake
in a previous version of this paper. We acknowledge financial
support from the project QAP (IST-FET FP6-015848) and from the Swiss
NCCR "Quantum Photonics" project.



\begin{thebibliography}{99}
\bibitem{distance} In Ref.~\cite{Weinfurter} the
authors suggest that it may be enough for the quantum state
of the atom to have collapsed, rather than for the result
of the measurement to have been recorded, which leads them
to a smaller required separation of 150 m. However, from
the strict perspective of testing the most general hidden
variable models this assumption is somewhat unsatisfactory.
\end{thebibliography}
\end{document}